\documentclass[final,5p,times,twocolumn,preprint]{elsarticle}
\usepackage{slashed}
\usepackage{eucal} 
\usepackage{bm}
\usepackage[hypertexnames=false,colorlinks=true,citecolor=blue,urlcolor=blue,
linkcolor=black]{hyperref}
\usepackage{amsmath,amssymb,amsfonts,latexsym}
\allowdisplaybreaks
\begin{document}
\begin{frontmatter}
\title{Tensor-polarized parton density in the $N \rightarrow \Delta$ transition}
\author{June-Young Kim}
\ead{jykim@jlab.org}
\author{Christian~Weiss}
\ead{weiss@jlab.org}
\address{Theory Center, Jefferson Lab, Newport News, VA 23606, USA}
\begin{abstract}
The generalized parton distributions for transitions between baryon states with different
masses have a forward limit in which they behave as parton densities
(light-front momentum transfer $\Delta^+, \Delta_T = 0$, energy transfer $\Delta^- \neq 0$).
These ``transition parton densities'' can realize spin/isospin quantum numbers not accessible
in the ground-state nucleon. The $N \rightarrow \Delta$ transition gives rise to a new
parton density proportional to the $1/2 \rightarrow 3/2$ spin transition tensor.
Its properties are derived, and its magnitude is estimated in the chiral
quark-soliton model based on the large-$N_c$ limit of QCD. 
\end{abstract}
\end{frontmatter}
\section{Introduction}
The generalized parton distributions (GPDs) describe the matrix elements of
QCD partonic operators between nucleon states with nonzero 4-momentum transfer
$\Delta \equiv p' - p \neq 0$. They represent the invariant form factors of the
partonic operators and contain extensive information on nucleon
structure \cite{Goeke:2001tz,Diehl:2003ny,Belitsky:2005qn,Boffi:2007yc}.
In the forward limit, $\Delta \rightarrow 0$, they reduce to the
conventional parton densities and can be interpreted as quark/gluon
momentum densities in the nucleon in light-front (LF) quantization.

The concept of GPDs can be extended to the transition matrix elements of the QCD partonic
operators between baryon states with different quantum numbers and masses, including
meson-baryon states ($\pi N, \pi\pi N, ...$) and resonances ($\Delta, N^\ast$)
\cite{Goeke:2001tz}. These ``transition GPDs'' present new opportunities for exploring
excited baryon structure in QCD and are an emerging area of theoretical and experimental
research \cite{Diehl:2024bmd}. In this context concepts such as QCD angular momentum
and the energy-momentum tensor form factors (mechanical properties) can be extended to
baryon resonances \cite{Kim:2023xvw,Kim:2022bwn,Alharazin:2023zzc}.

The transition GPDs between baryon states with different masses also have forward limit,
in which the LF momentum transfer between the states is zero, $\Delta^+ = 0$ and
$\bm{\Delta}_T = 0$, and only the LF energy transfer nonzero, $\Delta^- \neq 0$
(we use the standard definition of momentum and energy in LF quantization
\cite{Brodsky:1997de}). In this limit the transition GPDs represent matrix elements
of the quark/gluon momentum density between baryon states with the same momentum
but different energy. These ``transition parton densities'' (transition PDFs) can be
used to describe the structure of excited baryon states in QCD. Similar concepts
are used in nonrelativistic nuclear physics, where one considers matrix elements of
the particle density operator between nuclear states with the same 3-momentum (e.g.\ at rest)
but different energy.

Of particular interest are the $N \rightarrow \Delta$ transition GPDs \cite{Kim:2024hhd}.
The $1/2 \rightarrow 3/2$ spin transition has a rich structure with
vector and tensor characteristics \cite{Kim:2024hhd}.
The $N \rightarrow \Delta$ transitions can be analyzed systematically in the
large-$N_c$ limit of QCD, where $N \rightarrow N$ and $N \rightarrow \Delta$
transitions are connected through the dynamical spin-flavor symmetry
\cite{Witten:1979kh,Gervais:1983wq,Dashen:1993jt}. Experiments in
hard exclusive processes with $N \rightarrow \Delta$ transitions are performed
at JLab 12 GeV \cite{Diehl:2024bmd,CLAS:2023akb,Kroll:2022roq,Semenov-Tian-Shansky:2023bsy}.

In this work we consider the forward limit of the transition GPDs for general baryon states
and introduce the transition PDFs.
We then apply the concepts to the $N \rightarrow \Delta$ transition and show that there
is a unique transition PDF proportional to the $1/2 \rightarrow 3/2$
spin transition tensor. We derive its properties and estimate its magnitude
in the chiral quark-soliton model based on the large-$N_c$ limit of QCD.
\section{Transition GPDs}
The transition matrix element of the vector-type QCD partonic operator between general
baryon states is defined as 
\begin{align}
&\mathcal{M}_{B'B}
=\int \frac{d\lambda}{2\pi} e^{i\lambda x}
\langle B' | \bar{\psi} (-\lambda n /2) \slashed{n} \hat{T} \psi (\lambda n /2) | B  \rangle .
\label{matrix_element}
\end{align}
In the QCD operator, $\psi$ and $\bar\psi$ are the quark fields, $\hat{T} = \{1, \tau^a\}$
is a quark flavor matrix (two light flavors), $n$ is a light-like 4-vector, $n^2 = 0$,
and $\lambda$ is the parameter controlling the light-like separation of the fields.
The baryon states $|B\rangle \equiv |B(p, \sigma, i)\rangle$ and
$|B'\rangle \equiv |B'(p', \sigma', i')\rangle$ are characterized by their 4-momenta
$p$ and $p'$, spin projections $\sigma$ and $\sigma'$, and isospin projections $i$
and $i'$ (the preparation of the spin states is specified below).
The baryon 4-momenta are on mass shell, $p^2 = m^2$ and $p^{\prime 2} = m^{\prime 2}$,
and the baryons are regarded as a stable particles (the definition of matrix elements
in unstable states using complex analyticity is discussed in Ref.~\cite{Diehl:2024bmd}).
The average baryon 4-momentum and the 4-momentum transfer are defined as
\begin{align}
P \equiv (p' + p)/2, \hspace{2em} \Delta \equiv p' - p;
\label{P_Delta}
\end{align}
as result of mass shell conditions they satisfy
\begin{align}
&P\cdot \Delta = (m^{\prime 2} - m^2)/2,
\hspace{2em}
P^{2} + \Delta^{2}/4 = (m^{\prime 2} + m^2)/2.
\label{on_shell}
\end{align}
The light-like vector $n$ is chosen such that $n\cdot P = 1$.

The spin structure of the matrix element Eq.~(\ref{matrix_element}) is exhibited
by expanding it in bilinear forms in the baryon spin wave functions,
\begin{align}
&\mathcal{M}_{B'B} =
C_{\rm iso} (i', i)
\sum_{I} \bar{u}'(\sigma') \Gamma_I u(\sigma) \, H_{I}(x,\xi,t) .
\label{parametrization}
\end{align}
$u \equiv u(p, \sigma)$ and $u' \equiv u'(p', \sigma')$ are the
baryon spinors (the bispinor and vector indices are omitted for brevity), and
$\Gamma_I \equiv \Gamma_I (P, \Delta, n)$ are spinor matrices
formed from the 4-vectors $P, \Delta, n$ and invariant numerical
tensors. The number of structures is equal to the number of independent
spin components of the matrix element.
The functions $H_{I}(x,\xi,t)$ represent invariant amplitudes and
are referred to as transition GPDs. They depend on the partonic variable $x$
in Eq.~(\ref{matrix_element}), the skewness $\xi \equiv -\frac{1}{2} n\cdot \Delta$,
and the squared momentum transfer $t \equiv \Delta^{2}$.
$C_{\rm iso}(i', i)$ is an isospin factor, whose form depends on the flavor structure of
the QCD operator and the baryon isospin. 

The partonic interpretation of the matrix element Eq.~(\ref{matrix_element}) is
developed in LF quantization. LF 4-vector components are defined
as $p^\pm \equiv (p^0 \pm p^3)/\sqrt{2}, \bm{p}_T \equiv (p^1, p^2)$.
For a particle state with 4-momentum $p$ and mass $m$, $p^+$ and $\bm{p}_T$
are the LF momenta, and $p^- = (m^2 + |\bm{p}_T|^2)/(2 p^+)$ is the LF energy.
The light-like 4-vector $n$ is along the 3-direction with
$n^- = 1/P^+$ and $n^+ = 0$. The QCD partonic operator becomes the number
operator of quarks with LF momentum $k^+ = x P^+$.
The interpretation of the matrix element is the same as in the case of
$N \rightarrow N$ transitions, which has been discussed extensively in literature
\cite{Goeke:2001tz,Diehl:2003ny,Belitsky:2005qn,Boffi:2007yc}.
One distinguishes two regions: $|x| > |\xi|$, where the GPDs describe quark or
antiquark densities (DGLAP region); $0 < |x| < |\xi|$, where they describe
quark-antiquark distribution amplitudes (ERBL region).

The partonic interpretation of the matrix element Eq.~(\ref{matrix_element}) is
performed in the class of reference frames where $\bm{P}_T=0$
(vectors $n$ and $P$ collinear) and $P^+$ remains arbitrary (boost parameter).
The momentum transfer is specified by the LF plus and transverse components
\begin{align}
\Delta^+ = -2\xi P^+, \hspace{2em} \bm{\Delta}_T.
\end{align}
The minus components of $P$ and $\Delta$ are fixed by Eq.~(\ref{on_shell}) as
\begin{align}
P^{-} = \frac{ \xi(m^{\prime 2} - m^2)+ m^{\prime 2} + m^2 + |\bm{\Delta}_T|^{2}/2}{4P^{+}
(1-\xi^{2})},
\\
\Delta^{-} = \frac{ m^{\prime 2} - m^2 + \xi (m^{\prime 2} + m^2 + |\bm{\Delta}_T|^{2}/2)}
{2P^{+}(1-\xi^{2})}.
\end{align}
The spin states of the baryons are prepared as LF helicity states \cite{Brodsky:1997de}.
They are obtained from the rest-frame spin states by a sequence of LF boosts
(longitudinal + transverse) and transform kinematically under LF boosts.
The parton spin states obtained from the particle representation of the QCD
partonic operator are also LF helicity states. The spin structure of the
matrix element Eq.~(\ref{matrix_element}) in baryon and parton LF helicities
is subject to selection rules and permits a LF multipole expansion in the
transverse momentum transfer $\bm{\Delta}_T$ \cite{Diehl:2003ny,Kim:2024hhd}.
\section{Transition PDFs}
\label{sec:transition_pdf}
The interpretation of the transition matrix element Eq.~(\ref{matrix_element})
simplifies in the case of zero plus momentum transfer
\begin{align}
\xi = 0, \hspace{2em} \Delta^+ = 0 .
\label{xi_zero}
\end{align}
(i) The QCD operator can only sample quarks/antiquarks existing in the initial
baryon state, and leaves them with the same momentum in the final baryon state;
it cannot annihilate/create quark-antiquark pairs (only DGLAP region, no ERBL region).
The operator thus measures the quark/antiquark density in the baryonic system.
(ii) The baryon transition induced by the operator preserves the
LF plus momentum of the states and only transfers transverse momentum
and LF energy between the states,
\begin{align}
\bm{\Delta}_T \neq 0,
\hspace{2em}
\Delta^- = (m^{\prime 2} - m^2)/(2 P^+) \neq 0 .
\label{Delta_minus}
\end{align}
This is the standard representation of form factors in LF quantization (Drell-Yan-West frame)
\cite{Brodsky:1997de}. The transition matrix element can therefore be interpreted as the
transition form factor of the QCD operator, in analogy to nonrelativistic systems.
A nonzero $\Delta^-$ appears due to the mass difference but does not
affect the form factor interpretation.

Further, one can consider the limit that also the transverse momentum transfer
becomes zero, $\bm{\Delta}_T = 0$. In this situation the entire LF momentum transfer
($\Delta^+, \bm{\Delta}_T$) is zero, only the LF energy transfer $\Delta^-$ is
nonzero as in Eq.~(\ref{Delta_minus}), and also
\begin{align}
t = \Delta^2 = 2 \Delta^+\Delta^- - |\bm{\Delta}_T|^2 = 0.
\label{t_forward}
\end{align}
We refer to this as the ``forward limit'' of the transition matrix element.
Note that the 4-momentum transfer $\Delta$ does not fully vanish in this limit
if the baryon masses are unequal; rather is becomes a light-like
vector $\Delta \propto n$. Even so, the physical implications of the forward limit
are practically the same as in the case of equal baryon masses.
It corresponds to the situation where the operator transfers no LF momentum
but only LF energy to the system, i.e., excitation of the system at rest.

We define the ``transition PDFs'' associated with the $B \rightarrow B'$
transition as the forward limit of the transition GPDs,
\begin{align}
f_I(x) \equiv H_I(x, \xi = 0, t = 0).
\end{align}
They describe the quark/antiquark densities in the baryon as it undergoes
excitation at rest in the sense of LF quantization.
The interpretation depends on the partonic operator and the choice
of spin structures in the parametrization of the matrix element
Eq.~(\ref{parametrization}). The vector-type partonic operator in
Eq.~(\ref{matrix_element}) measures the unpolarized quark density,
with $f_I(x)$ describing the quark density at $x > 0$, and the
negative antiquark density at $x < 0$. The definitions can be
extended to the helicity-dependent or transversity-dependent
quark distributions.

An important constraint on the transition PDFs arises from the connection of the
partonic operator with the conserved vector current. The transition matrix element
of the vector current operator is defined as
\begin{align}
&\mathcal{J}_{B'B}^\mu
= \langle B' | \bar{\psi} (0) \gamma^\mu \hat{T} \psi (0) | B  \rangle,
\label{matrix_element_current}
\end{align}
and current conservation implies
\begin{align}
\Delta_\mu \mathcal{J}_{B'B}^\mu
=
\Delta^+ \mathcal{J}_{B'B}^-
+ \Delta^- \mathcal{J}_{B'B}^+
- \bm{\Delta}_T \cdot \bm{\mathcal{J}}_{B'B, T} = 0 .
\label{current_conservation}
\end{align}
In the forward limit $\Delta^+, \bm{\Delta}_T = 0$, and Eq.~(\ref{current_conservation})
reduces to
\begin{align}
\Delta^- \mathcal{J}_{B'B}^+ = 0.
\label{current_conservation_plus}
\end{align}
If the initial and final baryon mass are equal, $\Delta^- = 0$ by Eq.~(\ref{Delta_minus}),
and the equation is satisfied trivially. If the masses are unequal, $\Delta^- \neq 0$, and
current conservation requires $\mathcal{J}_{B'B}^+ = 0$, which is
\begin{align}
n_\mu \mathcal{J}_{B'B}^\mu = 0
\hspace{2em}
(\Delta^+, \bm{\Delta}_T = 0).
\end{align}
This contraction of the current is obtained when integrating the matrix element of
the partonic operator over $x$,
\begin{align}
\int_{-1}^1 dx \, \mathcal{M}_{B'B} = n_\mu \mathcal{J}_{B'B}^\mu = 0
\hspace{2em}
(\Delta^+, \bm{\Delta}_T = 0).
\label{sumrule_forward}
\end{align}
Current conservation thus implies a ``zero sum rule''
for the transition matrix element of the partonic operator in the forward limit.
It is specific to the transition between states with unequal masses and arises from
the constraint on the ``good'' component of the LF current, Eq.~(\ref{current_conservation_plus}).

When using the parametrization of the matrix element of the partonic operator,
Eq.~(\ref{parametrization}),
the sum rule Eq.~(\ref{sumrule_forward}) can be realized in two different ways
in the various terms: (i) The bilinear form can vanish in forward limit,
\begin{align}
\bar{u}' \Gamma_I u = 0
\hspace{2em}
(\Delta^+, \bm{\Delta}_T = 0).
\end{align}
In this case there is no restriction on the associated transition PDF.
(ii) The bilinear form can remain nonzero in forward limit,
\begin{align}
\bar{u}' \Gamma_I u \neq 0
\hspace{2em}
(\Delta^+, \bm{\Delta}_T = 0).
\end{align}
In this case the integral of the transition PDF must vanish,
\begin{align}
\int_{-1}^1 dx \, f_I(x) = 0.
\label{first_moment_zero}
\end{align}
\section{Rotationally symmetric representation}
The partonic interpretation of the transition matrix element Eq.~(\ref{matrix_element})
refers to the class of collinear frames with $\bm{P}_T = 0$ and arbitrary $P^+$.
The frames in this class are related by longitudinal boosts and specified by the choice of $P^+$.
A particular frame in this class corresponds to the 3D Breit frame with $\bm{P} = 0$,
where the ordinary 3-momenta of baryons are equal and opposite. This circumstance allows
one to ``match'' the LF structure in the collinear frames with the rotationally symmetric
structure in the Breit frame \cite{Kim:2023xvw}.

In the limit $\xi = 0$, Eq.~(\ref{xi_zero}), the 4-momenta $P$ and $\Delta$ in
the collinear frames have LF components
\begin{align}
&P^{+} = \text{arbitrary},
\hspace{2em}
P^- = \frac{ m^{\prime 2} + m^{2} - t/2}{4P^{+}},
\hspace{2em}
\bm{P}_T = 0,
\\
&\Delta^+ = 0,
\hspace{2em}
\Delta^- = \frac{ m^{\prime 2} - m^{2}}{2P^{+}},
\hspace{2em}
\bm{\Delta}_T \neq 0,
\end{align}
with $t = -|\bm{\Delta}_T|^2$. The particular choice
\begin{align}
P^+ = \frac{\sqrt{m^{\prime 2} + m^2 -t/2}}{2}
\end{align}
implies
$P^+ = P^-$ and thus $P^3 = (P^+ - P^-)/\sqrt{2} = 0$, so that $\bm{P} = (\bm{P}_T, P^3) = 0$.
This particular collinear frame is a 3D Breit frame. The ordinary 4-vector components
in this frame are
\begin{align}
& P^0 = \frac{\sqrt{m^{\prime 2} + m^2 - t/2}}{\sqrt{2}},
\hspace{2em}
\bm{P} = 0,
\label{P_symmetric}
\\
& \Delta^0 = -\Delta^3
= \frac{m^{\prime 2} - m^2}{\sqrt{2(m^{\prime 2} + m^2 - t/2)}},
\hspace{2em}
\bm{\Delta} = ({\bm\Delta}_T, \Delta^3).
\label{Delta_symmetric}
\end{align}
The initial and final baryon 3-momenta are equal and opposite
\begin{align}
&p = (E, -\bm{\Delta}/2),
\hspace{2em}
E = \sqrt{m^2 + |\bm{\Delta}|^2/4},
\\
&p' = (E', \bm{\Delta}/2),
\hspace{2em}
E' = \sqrt{m^{\prime 2} + |\bm{\Delta}|^2/4}.
\end{align}
It is important to note that $\Delta^3 \neq 0$ if $m' \neq m$, see Eq.~(\ref{Delta_symmetric}).
A nonzero longitudinal 3-momentum in the Breit frame is required to satisfy the condition
$\Delta^+ = 0$ for unequal baryon masses; this is so even for $t = -|\bm{\Delta}_T|^2 = 0$.
This highlights the difference between the baryons being at rest in the LF momentum
($\Delta^+ = 0$) and in ordinary momentum ($\Delta^3 = 0$).

In the Breit frame the baryon spin structure can be described by canonical spinors.
The matrix element then obeys constraints from 3D rotational symmetry, albeit with the
3-axis as preferred direction because of the light-like 4-vector.
The canonical spin structure in the Breit frame can be matched with the
LF helicity structure in the collinear frames (Melosh rotation).
The technique can be used in the analysis of spin
structure and the $1/N_c$ expansion of the matrix elements.
If $\bm{\Delta}_T = 0$, the motion of the initial/final baryon in the Breit frame
is along the 3-axis, and the LF helicity coincides with the canonical spin
projection on the 3-axis.

\section{$N \rightarrow \Delta$ tensor transition PDF}
We now apply the concepts to the GPDs and PDFs in the $N \rightarrow \Delta$ transition.
For definiteness we consider the $p \rightarrow \Delta^+$ transition matrix element
of the operator with $\hat{T} = \tau^3$,
\begin{align}
&\mathcal{M}_{\Delta^+p}
=\int \frac{d\lambda}{2\pi} e^{i\lambda x}
\langle \Delta^+ | \bar{\psi} (-\lambda n /2) \slashed{n} \tau^3 \psi (\lambda n /2)
| p  \rangle ,
\label{matrix_element_ndelta}
\end{align}
for which the isospin factor in Eq.~(\ref{parametrization})
is $C_{\rm iso} = \sqrt{2/3}$; the factors for the other isospin components
are given in Ref.~\cite{Kim:2024hhd}.

The spin structure of the $N \rightarrow \Delta$ transition matrix element
is analyzed in Ref.~\cite{Kim:2024hhd}. The spin-3/2 particle is described
by a vector-bispinor wave function $u^\alpha(\sigma')$. The expansion in
bilinear forms Eq.~(\ref{parametrization}) is given by
\begin{align}
&\mathcal{M}_{\Delta^+ p} =
\sqrt{\frac{2}{3}}
\sum_{I=M,E,C,X} \bar{u}_\alpha (\sigma') \mathcal{K}_I^{\alpha\mu} n_\mu
u(\sigma) \, H_{I}(x,\xi,t) .
\label{parametrization_ndelta}
\end{align}
The expressions of the bispinor matrices $\mathcal{K}_I^{\alpha\mu}$
are given in Ref.~\cite{Kim:2024hhd}.
The structures $I= M, E, C$ are connected with the magnetic, electric,
and Coulomb form factors in the $N \rightarrow \Delta$ transition matrix
element of the local vector current. The structure $I = X$ is unique to the
transition matrix element of the nonlocal partonic operator and given by
\begin{align}
\mathcal{K}_X^{\alpha\mu} n_\mu &= m_N n^\alpha \slashed{n} \gamma_{5} .
\label{K_X}
\end{align}
The bilinear form with this matrix remains nonzero in the forward limit
$\Delta^+, \bm{\Delta}_T = 0$. Evaluating it we obtain
\begin{align}
\bar{u}_{\alpha} (\sigma') [m_N n^{\alpha} \slashed{n}  \gamma_{5}] u (\sigma)
= \frac{2 m_N}{m_\Delta} Q^{33}(\sigma', \sigma) ,
\label{bilinear_quadrupole}
\end{align}
where $Q^{ij}$ is the $1/2 \rightarrow 3/2$ spin transition tensor
defined in \ref{app:tensor}, Eq.~(\ref{Q_cartesian}). The 33 component
appears from the contraction with the light-like 4-vector $n$.
The $N \rightarrow \Delta$ matrix element Eq.~(\ref{matrix_element_ndelta})
thus contains a spin tensor structure in the forward limit.
We define the tensor transition PDF as
\begin{align}
f_Q(x) \equiv H_X (x, \xi = 0, t = 0).
\end{align}
The structure is specific to the 1/2 $\rightarrow$ 3/2 spin transition of the nonlocal
partonic operator. It is absent in the 1/2 $\rightarrow$ 1/2 transition (no spin
tensor) and in the 1/2 $\rightarrow$ 3/2 transition of the
local vector current (no spatial direction from the light-like 4-vector).

Relations for the moments of $f_Q$ are obtained by integrating the matrix
element Eq.~(\ref{matrix_element_ndelta}) over $x$. The first moment is given
by the transition matrix element of the local vector current. Because the bilinear form
Eq.~(\ref{bilinear_quadrupole}) remains nonzero in the forward limit, current conservation
implies that the first moment of $f_Q$ is zero,
see Sec.~\ref{sec:transition_pdf} and Eq.~(\ref{first_moment_zero}),
\begin{align}
\int^{1}_{-1} dx \, f_Q(x) =  0.
\label{moment_first}
\end{align}
The second moment of the matrix element Eq.~(\ref{matrix_element_ndelta}) is given by the
transition matrix element of the rank-2 tensor operator
\begin{align}
\int^{1}_{-1} dx \, x \, \mathcal{M}_{\Delta^+ p}
&=  n_{\mu} n_{\nu} \,
\langle \Delta^{+}  | \bar{\psi}(0) \frac{i}{2}
\overleftrightarrow{\nabla}^{ \{\mu} \gamma^{\nu\} }
\tau^3 \psi(0) |  p \rangle
\label{moment_second}
\\
&= \sqrt{\frac{2}{3}} \, n_{\mu} n_{\nu} \, \bar{u}_{\alpha} 
\left[m_N \gamma^{ \{\mu} g^{\nu \} \alpha}  \gamma_{5} \right] u \, F_{4}(t)   + ...
\end{align}
where $\nabla$ is the QCD covariant derivative with
$\overleftrightarrow{\nabla}\equiv (\overrightarrow{\nabla} - \overleftarrow{\nabla})/2$,
$\{\mu \nu\} = \mu\nu + \nu \mu$, and $F_4$ is the form
factor in the parametrization of the matrix element
introduced in Ref.~\cite{Kim:2022bwn}. One obtains
\begin{align}
\int^{1}_{-1} dx \, x \, f_Q(x) = 2 F_{4}(t).
\end{align}
The tensor operator here can be regarded as the
isovector version of the quark part of the QCD energy-momentum tensor;
this operator is not a conserved current. 
\section{Large-$N_c$ estimate}
We now compute the spin tensor transition PDF
using the chiral quark-soliton model of baryons \cite{Diakonov:1987ty,Christov:1995vm}.
The model is based on the large-$N_c$
limit of QCD and effective dynamics resulting from the spontaneous breaking of chiral symmetry.
It has been successfully applied to $N \rightarrow N$ PDFs
\cite{Diakonov:1996sr,Diakonov:1997vc,Pobylitsa:1998tk}
and GPDs \cite{Petrov:1998kf,Penttinen:1999th,Ossmann:2004bp}
and has provided insights into their general properties.
It also connects $N \rightarrow N$ and $N \rightarrow \Delta$ matrix elements
through the dynamical symmetry
in the large-$N_c$ limit \cite{Goeke:2001tz}.

The calculation is performed in the rotationally symmetric representation of the
transition matrix element in the Breit frame, defined by
Eqs.~(\ref{P_symmetric}) and (\ref{Delta_symmetric}) with $\bm{\Delta}_T = 0$.
With $m_{\Delta, N} = \mathcal{O}(N_c)$ and $m_\Delta - m_N = \mathcal{O}(N_c^{-1})$
one has $\Delta^{3} = \mathcal{O}(N_c^{-1})$, i.e.\ the 3-momentum transfer required
by the baryon mass difference is small, and the calculation can be performed with
the baryons at rest, $\Delta^3 = 0$.

The dynamics of the chiral quark-soliton model and the calculation of matrix elements
of partonic operators are described in detail in Refs.~\cite{Diakonov:1996sr,Diakonov:1997vc}.
Baryons are characterized by a classical chiral field of ``hedgehog'' form
(isospin aligned with the spatial direction). Baryon states with spin-isospin quantum numbers
appear from the quantization of the (iso-) rotational zero modes.
The partonic QCD operator is first averaged in the mean-field state of quarks in the
background of the classical chiral field. The mean field state is then subjected to
collective (iso) rotations, and the operator average is expanded in powers of the angular
velocity $\Omega = \mathcal{O}(N_c^{-1})$. The rotations are then quantized,
and the mean-field average of the operator is projected on the baryon spin-isospin states.

In the case of the isovector unpolarized partonic operator of Eq.~(\ref{matrix_element_ndelta}),
the mean-field average is nonzero only at first order in $\Omega = \mathcal{O}(N_c^{-1})$;
the matrix elements are therefore
subleading in the $1/N_c$ expansion. The matrix element between
$N$ and $\Delta$ states is obtained as
\begin{align}
\mathcal{M}_{\Delta^+p} = - \frac{2\sqrt{2}}{I} \,
\langle \{ \hat{D}^{33}, \hat{S}^{3} \} \rangle_{\Delta^+ p}
\, F_{\mathrm{mf}}(x) + ...
\label{ndelta_largenc}
\end{align}
Here $I = \mathcal{O}(N_c)$ is the moment of inertia of the mean field,
$\hat{D}^{33}$ and $\hat{S}^{3}$ are operators in the collective rotation
degrees of freedom (see e.g.\ Ref.~\cite{Pobylitsa:1998tk}),
$\langle...\rangle_{\Delta^+ p}$ is the matrix element between the
$p$ and $\Delta^+$ rotational wave functions,
and $F_{\mathrm{mf}}(x)$ is the pertinent mean-field parton density.
In Eq.~(\ref{ndelta_largenc}) ... denote structures that do not
contribute to the spin tensor transition.
The matrix element between rotational states evaluates to
\begin{align}
\langle \{ \hat{D}^{33}, \hat{S}^{3} \} \rangle_{\Delta^+ p}
&= -\frac{ 1}{\sqrt{3}} Q^{33} (\sigma', \sigma), 
\end{align}
where $\sigma$ and $\sigma'$ are the $N$ and $\Delta$ spin projections.
Equating the large-$N_c$ matrix element with the corresponding structure in
the parametrization Eqs.~(\ref{parametrization_ndelta}) and (\ref{bilinear_quadrupole}),
and using the simplifications from the large-$N_c$ kinematics ($m_N = m_\Delta$),
we obtain the tensor transition PDF as
\begin{align}
f_Q(x) = \frac{F_{\rm mf}(x)}{I} .
\label{largenc_general}
\end{align}
The inverse moment of inertia is proportional to the $N$-$\Delta$ mass splitting
and can be inferred from it,
\begin{align}
m_\Delta - m_N = \frac{3}{2I}.
\label{inertia_ndelta}
\end{align}

To evaluate the mean-field parton density in Eq.~(\ref{largenc_general}) we employ 
an expansion in derivatives of the classical chiral field sustaining the mean-field state
(gradient expansion) \cite{Diakonov:1996sr}.
The technique has been used for numerical computations of PDFs and GPDs and preserves
the partonic sum rules resulting from the connection with local operators
\cite{Diakonov:1996sr,Petrov:1998kf,Penttinen:1999th,Ossmann:2004bp}. We obtain
\begin{align}
f_Q(x) &= \frac{m_N f^{2}_{\pi}}{4\sqrt{2} I}
\int \frac{d^{3}k}{(2\pi)^{3}} \;
\text{sign}(x) \, \Theta\left[ x \left(\frac{k^3}{m_N} - x\right)\right] \, \frac{1}{k^{3}} 
\nonumber \\
&\times \left( 1 - \frac{3 (k^3)^2}{|\bm{k}|^2} \right) \, \Pi^{2}(\bm{k}).
\label{gradient}
\end{align}
Here $f_\pi$ is the pion decay constant. The integral is over the 3-momentum of
chiral field. $\Pi(\bm{k})$ is the radial Fourier transform of the chiral field
in position space ($r \equiv |\bm{x}|, \hat{\bm{x}} \equiv \bm{x}/r$),
\begin{align}
&U(\bm{x}) \equiv \exp [i \hat{\bm{x}}\bm{\tau} P(r) ] = \cos P(r) + i\hat{\bm{x}}\bm{\tau}
\sin P(r),
\\
& \Pi(\bm{k}) \equiv 4 \pi \int dr \, r^{2} j_{1}(r |\bm{k}|) \sin P(r).
\end{align}
The radial profile function satisfies $P(0) = -\pi$ and $P(r \rightarrow \infty) = 0$.
The angular structure in the integral Eq.~(\ref{gradient}) is specific to the
spin tensor transition ($L = 2$).

The gradient expansion result Eq.~(\ref{gradient}) satisfies the sum rule
Eq.~(\ref{moment_first}). Integrating Eq.~(\ref{gradient}) over $x$, we get
\begin{align}
\int^{\infty}_{-\infty} dx \,
\text{sign}(x) \, \Theta\left[ x \left(\frac{k^3}{m_N} - x\right)\right]
= \frac{k^3}{m_N},
\end{align}
which cancels the factor $1/k^3$ in the integrand of Eq.~(\ref{gradient}).
The integral of Eq.~(\ref{gradient}) then becomes rotationally symmetric and
vanishes because $3(k^3)^2 = |\bm{k}|^2$ after angular averaging.
When integrated over $x$, the matrix element
has no knowledge of the spatial direction of light-like 4-vector $n$.

\begin{figure}[t]
\centering
\includegraphics[width=0.9\columnwidth]{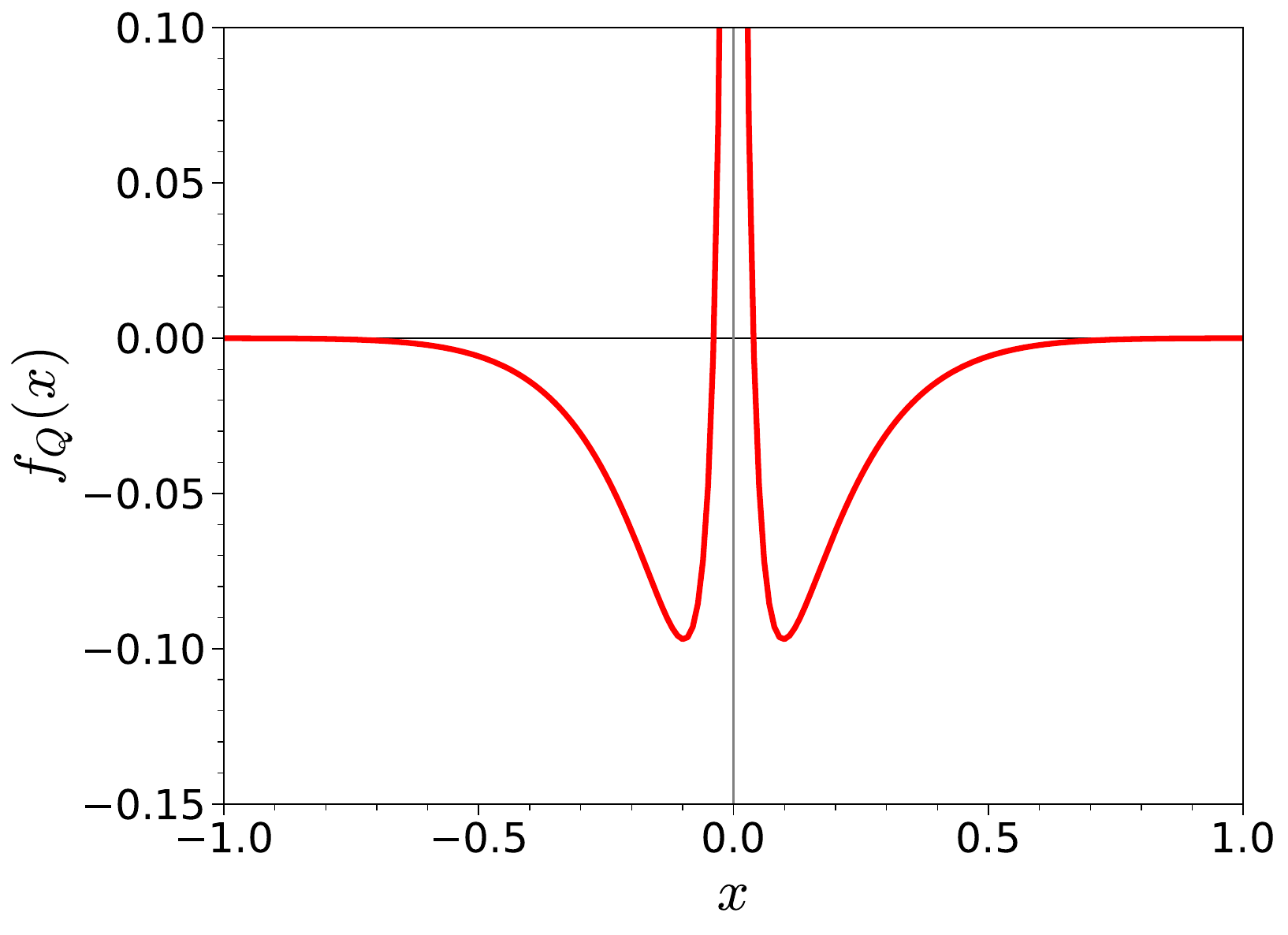}
\caption{Numerical estimate of the $N \rightarrow \Delta$ tensor transition PDF
$f_Q(x)$ obtained in the chiral quark-soliton model with the gradient expansion;
see Eqs.~(\ref{inertia_ndelta}) and (\ref{gradient}).}
\label{fig:quadrupole}
\end{figure}

Figure~\ref{fig:quadrupole} shows the numerical result for the tensor transition PDF
in the chiral quark-soliton model obtained from Eqs.~(\ref{inertia_ndelta}) and (\ref{gradient}).
The distribution is even in $x$. One observes that (for $x > 0$)
the function changes sign at $x \sim 0.1$. This property follows from the angular dependence of the
integrand in Eq.~(\ref{gradient}); the sign change is required for the first moment
to vanish according to the sum rule Eq.~(\ref{moment_first}).
The magnitude of the tensor transition PDF is about an order-of-magnitude smaller
than that of standard PDFs with sum rules $\int dx \, f(x) = \mathcal{O}(1)$.
%
%

It is instructive to compare the spin tensor PDF in
$N \rightarrow \Delta$ transitions with the ``spin vector'' PDF in $N \rightarrow N$ transitions.
The latter is defined as the forward limit of the nucleon GPDs,
\begin{align}
f_D(x) \equiv [H^{u+d} + E^{u+d}] (x, \xi = 0, t=0).
\label{f_D}
\end{align}
The sum of Dirac and Pauli GPDs corresponds to the dipole structure in the
multipole expansion of the matrix element; this combination
of GPDs enters in the nucleon spin sum rule \cite{Ji:1996ek,Polyakov:2002yz}.
The flavor-singlet vector PDF in Eq.~(\ref{f_D}) is subleading in $1/N_c$
and thus has the same status as the flavor-nonsinglet tensor PDF.
In the chiral quark-soliton model both distributions appear at $\mathcal{O}(\Omega)$
and have similar properties. In the gradient expansion the vector distribution
is obtained as \cite{Ossmann:2004bp}
\begin{align}
f_D(x) &=  
\frac{m^{2}_N f^{2}_{\pi}}{2 I}
\int \frac{d^{3}k}{(2\pi)^{3}} \;
\text{sign}(x) \, \Theta\left[ x \left(\frac{k^3}{m_N} - x\right)\right] \, \frac{1}{k^{3}} 
\nonumber \\
&\times k^3 \, \frac{\Pi^{2}(\bm{k})}{|\bm{k}|^{2}} .
\label{gradient_dipole}
\end{align}
This result has the same form as Eq.~(\ref{gradient}) for $f_Q$, only the angular structure
is replaced by $k^3$ ($L = 1$). It shows the close connection between the ``dipole'' and
``quadrupole'' structures emerging from the mean-field picture at large $N_c$.
(The connection
between $f_Q$ and $f_D$ discussed here is not a group-theoretical large-$N_c$ relation
resulting from spin-flavor symmetry, but a dynamical connection
appearing in the mean-field picture. The two distributions arise as different angular
averages of the partonic operator in the same mean-field state.)

The numerical result for the $N \rightarrow \Delta$ spin tensor PDF shows a
rise of the function for $x \rightarrow 0$ (see Fig.~\ref{fig:quadrupole}).
The present mean-field calculation assumes ``nonexceptional''
$x = \mathcal{O}(N_c^{-1})$ and cannot predict the $x \rightarrow 0$ behavior of the PDF.
The local behavior of the distribution at $x \rightarrow 0$ should therefore not be
regarded as a prediction of the model, even though the integral behavior correctly
reproduces the sum rule for the first moment.\footnote{The behavior of the PDF at
$x \rightarrow 0$ is related to the limit of large spatial separation of the fields
in the QCD operator Eq.~(\ref{matrix_element}).
In this regime the rigid rotor quantization of the collective (iso) rotations
becomes inadequate and produces artifacts. Such effects have been observed
in studies of the interplay of the chiral and large-$N_c$ limits in
soliton models \cite{Cohen:1996zz}.}

\section{Summary and extensions}
The results of the present study are: (i)~A natural concept of transition PDFs
emerges from the transition GPDs in the limit $\Delta^+, \bm{\Delta}_T = 0$.
The partonic operator measures a quark/antiquark density; the matrix element
involves zero LF momentum transfer, only LF energy transfer,
and describes a property of the system at rest in LF momentum.
(ii)~In the corresponding Breit frame representation, the baryons are
not at rest in ordinary 3-momentum but move with momenta proportional
to the baryon mass difference.
(iii)~The $N \rightarrow \Delta$ transition gives rise to a spin tensor PDF.
The structure is unique to the $1/2 \rightarrow 3/2$ spin transition and results from
the projection of the spin transition tensor on the spatial direction provided
by the partonic operator.
(iv)~The tensor transition PDF is subleading in the $1/N_c$ expansion.
(v)~The chiral quark-soliton model predicts that the tensor transition PDF is
nonzero but numerically small, at least an order-of-magnitude smaller than regular nucleon PDFs.

The investigations could be extended several directions.
The present study considers only the spin tensor component of the
$N \rightarrow \Delta$ transition matrix element that remains present
at $\bm{\Delta}_T = 0$ (monopole in $\bm{\Delta}_T$).
The $N \rightarrow \Delta$ transition matrix element has higher orbital multipoles
(dipole and quadrupole in $\bm{\Delta}_T$), which are accompanied by other spin
structures \cite{Kim:2024hhd}.
Extending the analysis to these structures requires keeping $\bm{\Delta}_T \neq 0$
so as to separate the various orbital multipoles.

From the transition GPDs at $\Delta^+ = 0$ and $\bm{\Delta}_T \neq 0$ one can define
the impact-parameter-dependent parton distribution as the Fourier transform
$\bm{\Delta}_T \rightarrow \bm{b}$ and quantify the transverse spatial distribution
of the partons in the transition PDF \cite{Burkardt:2000za,Burkardt:2002hr}.
This opens new possibilities for exploring the spatial structure
of baryon resonances in QCD \cite{Diehl:2024bmd}.

The analysis could be extended to the quark helicity densities measured by the partonic
operators with spinor matrix $\slashed{n}\gamma_5$. It could also be extended to the
quark transversity operators with the chiral-odd spinor matrix $n_\alpha \sigma^{\alpha T}$.
Chiral-odd transition GPDs are sampled in exclusive pion production with $N \rightarrow \Delta$
transitions, if the reaction is described using QCD factorization with the chiral-odd
pion distribution amplitude \cite{Kroll:2022roq}. Chiral-odd $N \rightarrow N$
GPDs have been analyzed in the large-$N_c$ limit \cite{Schweitzer:2016jmd,Tezgin:2024tfh,
Kim:2024ibz,Kim:2025mol}.

The question of possible measurements of the tensor polarized $N \rightarrow \Delta$
transition GPDs in hard exclusive processes (deeply-virtual Compton scattering
\cite{Semenov-Tian-Shansky:2023bsy}, meson production \cite{Kroll:2022roq})
should be addressed in a future study. This includes identifying the experimental polarization
conditions needed to isolate the tensor-polarized structure (target polarization,
analysis of the $\Delta$ decay, recoil polarization of decay nucleon).

The $1/2 \rightarrow 3/2$ tensor transition PDF described here is similar to the tensor-polarized
PDF in spin-1 hadrons such as the deuteron (corresponding to the $1 \rightarrow 1$ transition),
which can be measured in deep-inelastic scattering on a tensor-polarized
target \cite{Frankfurt:1983qs,Hoodbhoy:1988am,Poudel:2025nof}.
It would be interesting to explore the analogy between the two structures and their
possible measurement.
\appendix
\section{Spin transition vector and tensor}
\label{app:tensor}
The $1/2 \rightarrow 3/2$ spin transition is characterized by a transition vector
and tensor, formed from the initial and final spin wave functions. Their 3D cartesian
components are defined as
\begin{align}
V^{i}(\sigma', \sigma)&\equiv  \sqrt{\frac{3}{2}} \sum_{\lambda' \tau'}
C^{\frac32 \sigma' }_{1 \lambda' \frac12 \tau'} \delta_{\tau' \sigma} \epsilon^{*i}_{\lambda'} ,
\\
Q^{ij}(\sigma', \sigma)&\equiv 
\sum_{\lambda' \tau'} C^{\frac32 \sigma' }_{1 \lambda' \frac12 \tau'}  \frac{1}{2}
\sigma^{ \{i}_{\tau' \sigma} \epsilon^{* j\} }_{\lambda'},
\label{Q_cartesian}
\end{align}
where $\epsilon^{i}_{\lambda'}$ are the spin-1 3-vector wave functions, $\sigma^i$ 
are the Pauli matrices, and $\{ij\} \equiv ij + ji$ \cite{Varshalovich:1988ifq}.
The tensor is symmetric and traceless, $Q^{ii} = 0$.
The spherical components are defined using the general prescription
\begin{align}
T_{LM}(\sigma',\sigma) \equiv \frac{\sqrt{2L+1}}{\sqrt{2S+1}} C^{S' \sigma'}_{S\sigma L M}
\hspace{2em} (L=1,2,..),
\end{align}
which generalizes the standard definition to the case of non-diagonal spin transitions
($S' \neq S$). Thus
\begin{align}
V_{1M}(\sigma',\sigma) = \sqrt{\frac{3}{2}} C^{\frac{3}{2} \sigma'}_{\frac{1}{2} \sigma 1 M},
\\
Q_{2M}(\sigma',\sigma) = \sqrt{\frac{5}{2}} C^{\frac{3}{2} \sigma'}_{\frac{1}{2} \sigma 2 M}.
\end{align}
The cartesian and spherical components are connected by the standard relations
\cite{Varshalovich:1988ifq}. Specifically,
\begin{align}
Q^{33}(\sigma', \sigma)
\; = \; \sqrt{\frac{2}{3}} \, Q_{20}
\; = \; \sqrt{\frac{2}{3}} \times
\left(\begin{array}{r r} 0 & 0 \\[0ex] 1 & 0 \\[0ex] 0 & -1 \\[0ex] 0 & 0
\end{array}\right)_{\sigma'\sigma} .
\end{align}
The transition vector and tensor defined here generalize the concept of polarization vector
and tensor from the diagonal $1 \rightarrow 1$ to the non-diagonal
$1/2 \rightarrow 3/2$ spin transition.

Acknowledgments:
This material is based upon work supported by the U.S.~Department of Energy, Office of Science,
Office of Nuclear Physics under contract DE-AC05-06OR23177. 

The research reported here is connected with the Topical Collaboration ``3D quark-gluon
structure of hadrons: mass, spin, tomography'' (Quark-Gluon Tomography Collaboration) supported by
the U.S.~Department of Energy, Office of Science, Office of Nuclear Physics under
contract DE-SC0023646.

\bibliography{NDeltaPDFs_letter}

\end{document}